\documentclass[aps,10pt,prd,groupedaddress,preprint]{revtex4-1}
\usepackage[utf8]{inputenc}
\usepackage[T1]{fontenc}
\usepackage{hyperref}
\usepackage{xcolor}
\usepackage{graphicx}
\usepackage{amsmath,amssymb}
\usepackage{bm}

\newcommand{\pa}{\partial}
\newcommand{\ma}[1]{{\mathrm{#1}}}

\begin{document}
	
\title{\boldmath Cosmological disformal invariance}

\author{Guillem Dom\`enech}
\email{guillem.domenech{}@{}yukawa.kyoto-u.ac.jp}
\affiliation{Yukawa Institute for Theoretical Physics, 
	Kyoto University, Kyoto 606-8502, Japan}
\author{Atsushi Naruko}
\email{naruko{}@{}th.phys.titech.ac.jp}
\affiliation{Department of Physics, Tokyo Institute of Technology,
	2-12-1 Ookayama, Meguro-ku, Tokyo
	152-8551, Japan}
\author{Misao Sasaki}
\email{misao{}@{}yukawa.kyoto-u.ac.jp}
\affiliation{Yukawa Institute for Theoretical Physics, 
	Kyoto University, Kyoto 606-8502, Japan}


\preprint{YITP-15-22}

\begin{abstract}
	The invariance of physical observables under disformal transformations is
	considered. 
	It is known that conformal transformations leave physical observables
	invariant. 
	However, whether it is true for disformal transformations is still
	an open question.
	In this paper, it is shown that a pure disformal transformation
	without any conformal factor
	is equivalent to rescaling the time coordinate.
	Since this rescaling applies equally to all the physical quantities,
	physics must be invariant under a disformal transformation,
	that is, neither causal structure, propagation speed nor any other property of
	the fields are affected by a disformal transformation itself.
	This fact is presented at the action level for gravitational and matter fields
	and it is illustrated with some examples of observable quantities.
	We also find the physical invariance for cosmological perturbations 
	at linear and high orders in perturbation, extending previous studies.
	Finally, a comparison with Horndeski and beyond Horndeski theories
	under a disformal transformation is made.
\end{abstract}



\maketitle
\flushbottom
\section{Introduction \label{intro}}
Observational cosmology reveals how small our knowledge of the Universe is. 
For example, two of the main unsolved mysteries are periods of accelerated 
expansion of the Universe, inflation and dark energy, whose underlying sources 
remain unknown. Plenty of theoretical models (for an extensive review
 see \citep{clifton2012modified}) point towards a scalar field as the main 
responsible for such an accelerated expansion. Interestingly, despite the 
good agreement of the standard $\Lambda$CDM model with current 
data \citep{ade2015planck2,ade2015planck3}, some tension appears from direct 
measurements of the time evolution of the Hubble parameter \citep{sahni2014model}. 
Furthermore, a recent release of the CMB observation by Planck~\citep{ade2015planck},
which supports inflation, seems to favour those inflationary models with
a non-minimal coupling, a.k.a. the scalar-tensor theory of 
gravity \cite{brans1961mach,fujii2003scalar,faraoni2004cosmology} that 
includes the so-called $f(R)$ theory \citep{whitt1984fourth,jakubiec1988theories}.

Conformal transformations have been proved to play a very important role in 
physics \citep{kastrup2008advancements}. In particular, in the scalar-tensor 
theory they are not only a useful mathematical tool but they leave observational
physics invariant at a classical level (for a discussion supporting this point of view
see \cite{makino1991density, faraoni2007pseudo, deruelle2011conformal, gong2011conformal, white2012curvature, catena2007einstein,chiba2013conformal,chiba2008extended,jarv2014invariant,domenech2015conformal}; also see \citep{capozziello2006cosmological,nojiri2006modified,briscese2007phantom,nojiri2011unified} for a discussion in a different point of view in the context of $F(R)$ gravity). 
Consequently, the notion of conformally related frames naturally appears. 
The Jordan frame or matter frame, where the scalar field is non-minimally coupled 
to the metric $\tilde{g}_{\mu\nu}$ but matter is minimally coupled, and 
the Einstein frame or gravitational frame, where the scalar field is minimally 
coupled to the metric $g_{\mu\nu}$ but matter is dilatonically coupled to 
the scalar field. 
An alternative way to understand conformal transformations in the 
scalar-tensor theory is the following. One has a gravitational 
metric $g_{\mu\nu}$ which satisfies the Einstein equations and a matter 
metric related to the former by a field dependent rescaling, 
say $\tilde{g}_{\mu\nu}=\Omega(\phi)^2g_{\mu\nu}$.

From this point of view, one may wonder whether there are more general 
transformations that lead to a new form of the matter metric. In other words, 
is there any other way that matter can non-trivially couple to gravity 
through a scalar field? This question was studied by Bekenstein 
in \citep{bekenstein1993relation} where a new class of transformations, 
called disformal transformations, were proposed. The idea behind such 
transformations is that matter is coupled to a metric which is not just 
a rescaling of the gravitational metric but it is stretched in a particular 
direction, given by the gradient of a scalar field.

Disformal transformations can be motivated from brane world models and 
from massive gravity theories (see \citep{brax2012shining,zumalacarregui2013dbi} 
and references therein)
and have been applied to inflation \citep{kaloper2004disformal}, dark
energy \citep{koivisto2008disformal,zumalacarregui2010disformal,van2015disformal},
varying speed of light 
models \citep{clayton2001scalar,moffat2003bimetric,magueijo2003new,
magueijo2009bimetric}, atomic physics \citep{brax2014explaining} 
and mimetic gravity \citep{deruelle2014disformal}.
Consequently, considerable effort has been made to look for constraints and how 
to avoid them, e.g. screening mechanisms, for disformally coupled matter 
models \citep{brax2012shining,zumalacarregui2010disformal,van2015disformal,
sakstein2014disformal,sakstein2015towards,koivisto2012screening,
van2013constraints,brax2013cosmological,brax2014constraining,zumalacarregui2013dbi}. 
In addition, with the recent rediscovery of the Horndeski 
theory \citep{horndeski1974second,Deffayet:2011gz,kobayashi2011generalized}, 
which is the general scalar-tensor theory with second order field equations 
of motion, and its generalization, known as beyond Horndeski or GLPV 
theory \citep{gleyzes2013essential,gleyzes2014healthy,gleyzes2014exploring,
gleyzes2014unifying} (see also its further extension \cite{Gao:2014soa}), 
attention has been paid to its mathematical invariance under disformal 
transformations in those 
theories \citep{zumalacarregui2014transforming,bettoni2013disformal}. 
If observational physics turns out to be invariant under disformal transformations 
that will provide us with a powerful mathematical and physical tool to 
classify or work within the Horndeski theory (for 
example \citep{zumalacarregui2013dbi,bettoni2015shaken}), and to make progress
in our understanding of the symmetries of gravity. In this 
direction, refs.~\citep{creminelli2014resilience,minamitsuji2014disformal,
tsujikawa2014disformal} study the disformal invariance of cosmological 
perturbations at the linear level with a positive answer. For a recent development see references \citep{watanabe2015multi,motohashi2015disformal,gleyzes2015effective}. In particular, see \citep{watanabe2015multi} for a multi-field extension of disformal transformations.

That being said, a note is in order. 
It is generally believed that physics does not change under a non-singular field redefinition. However, a field redefinition of a metric is more subtle when it comes to interpretations. A simple example would be a conformal transformation which could lead us from an homogeneous and isotropic expanding universe to a static spacetime (see Deruelle \& Sasaki \cite{deruelle2011conformal}). The invariance of physically observable quantities before and after the transformation is not immediately clear. Therefore, and also to clearly understand how to interpret the results, one has to \textit{explicitly} check the invariance.

Turning to the case of a disformal transformation, 
it is no longer just a simple 
field redefinition, because it involves derivatives of the scalar field.
Thus there may be a higher level of subtlety than in the case of a conformal transformation.
 For this reason, we dedicate this work to study the effects of a pure disformal 
transformation, its interpretations and the invariance of physical observables 
in cosmology.

For the sake of simplicity and clarity, we restrict 
ourselves to work in the comoving, or the uniform $\phi$ slicing,
 on which the scalar field is homogeneous. 
In this sense, we discuss \textit{cosmological} disformal transformations
by implicitly assuming the existence of $\phi$-constant spacelike hyper-surfaces,
 which is a common and reasonable assumption in cosmology. 
We show that a pure cosmological
disformal transformation without any conformal factor is equivalent 
to rescaling the time coordinate, which could even be regarded as a time 
coordinate transformation. This is checked at the action level for 
gravitational and matter fields, which ensures that physics is invariant under 
disformal transformations. Consequently, one can work in the frame one 
considers most suitable for either computations or interpretations.

The paper is organized as follows. In section \ref{sec:pure} we briefly review 
disformal transformations and we focus on pure disformal transformations, which 
are intuitively shown to be equivalent to rescaling the time coordinate when 
applied to cosmology. Afterwards, in section \ref{sec:action} we proceed to check 
that this is the case for gravitational and matter fields at the action level, 
with a comparison to Horndeski theory. Let us stress that throughout 
sections \ref{sec:pure} and \ref{sec:action} we focus on the effects of a 
disformal transformation \textit{itself}, i.e. no comparison between fields, c.f. 
scalar and matter fields is made. Afterwards, in section \ref{sec:implications} 
we explicitly compute the disformal invariance of some observable quantities 
considering the system as a whole, i.e. gravitational and matter sector 
altogether, and we emphasise on the choice of frames and its invariance.  
Finally, in section \ref{sec:conclusions} we summarize our results.

\section{Pure disformal transformation}
\label{sec:pure}

The general form of a disformal transformation is 
given by \citep{bekenstein1993relation}
\begin{align}
\bar{g}_{\mu\nu}=G(\phi,X)g_{\mu\nu}+F(\phi,X)\phi_{,\mu}\phi_{,\nu},
\label{eq:dis}
\end{align}
where $X=-\frac{1}{2}g^{\mu\nu}\phi_{,\mu}\phi_{,\nu}$ and 
$\phi_{,\mu}\equiv\partial_\mu \phi$. The first term in the right hand side
 corresponds to a conformal transformation, i.e. a rescaling of the metric, 
whereas the latter is a pure disformal transformation, i.e. the metric is 
stretched in the direction of $\phi_{,\mu}$. In this work, we restrict 
ourselves to the case of a pure disformal transformation, that is 
$G(\phi,X)=1$, as any conformal transformation can be \textit{generally} 
done afterwards.

In cosmology it is a reasonable assumption that $\phi_{,\mu}$ is regular and
timelike everywhere, and hence one can choose the comoving, or the uniform 
$\phi$ slicing on which the scalar field is homogeneous. 
Throughout this paper, we take the comoving slicing to simplify
 the argument and make clear physical interpretations and consequences.
In fact one immediately notices from \eqref{eq:dis} that
 under this pure cosmological disformal transformation
 only the time-component of the metric, namely the lapse function, is modified.
To see this explicitly, let us express a pure ``disformation''
in the form of the $(3+1)$-decomposition,
\begin{align}
&g\equiv ds^2=-N^2dt^2+g_{ij}(dx^i+N^i dt)(dx^j+N^j dt)
\cr
\to\quad&
\bar{g}\equiv
d\bar{s}^2=-\bar{N}^2dt^2+\bar{g}_{ij}(dx^i+\bar{N}^i dt)(dx^j+\bar{N}^j dt)\,,
\end{align}
where
\begin{align}
\bar{N}^2=N^2\alpha^2\,,\quad \bar{N}^i=N^i\,,
\quad \bar{g}_{ij}=g_{ij}\,,
\label{eq:disdown}\   
\end{align}
where for the sake of simplicity we defined
\begin{align}
\alpha^2=1-2F(\phi,X)X.
\label{eq:alpha}
\end{align}
It should be noted that now $X$ reduces to $- (1/2) g^{t t} \dot{\phi}^2$
 where the scalar field only depends on the time coordinate 
 but $g^{t t}=-N^{-2}$ has spatial dependence in general, which implies 
 the spatial dependence of the disformal factor, $F$, as well as of $\alpha$.
From \eqref{eq:disdown} we infer that in order to preserve the Lorentzian 
signature of the metric one must require 
$\alpha^2>0$ \citep{bekenstein1993relation}, which is assumed throughout 
this paper. Due to the fact that just a particular component of the metric 
is modified, one may naively think that the causal structure and the 
propagation speed are altered as well. 
In fact, it is rather the opposite as shown later.

A note is in order. Deruelle and Rua \citep{deruelle2014disformal} 
showed that a disformal transformation 
does not alter Einstein's equations in general, although the form of such is quite different when expressed in terms of the transformed metric. One may naively expect this result due to the fact that a disformal transformation is a 
re-parametrization of the metric.

However, there is a particular class of
 disformal transformations which leads to a departure from usual General 
Relativity (GR). In fact, its effect is to give rise to a source term in 
Einstein's equations even in the absence of matter fields. This particular 
class includes the Mimetic Dark Matter 
model \citep{chamseddine2014cosmology,chamseddine2013mimetic} where this 
new source term imitates a dark matter component. Concretely, the mimetic 
gravity condition gives a constraint between the conformal and disformal 
factors. In our notation it reads
\begin{align}
F(\phi,X)=\frac{1}{2X}G(\phi,X)-f(\phi),
\end{align}
where $f(\phi)$ is an arbitrary positive function of
 the scalar field alone. 
For a pure disformal transformation (G=1) it reduces to
\begin{align}
F(\phi,X)=\frac{1}{2X}-f(\phi),
\end{align}
which plugged back in \eqref{eq:disdown} leads us to a degenerate metric 
without lapse, that is
\begin{align}
d\bar{s}^2=- f(\phi)(\pa_t\phi)^2 dt^2
+\bar{g}_{ij}(dx^i+\bar{N}^i dt)(dx^j+\bar{N}^j dt)\,.
\end{align}
Note that there will no longer be any constraint equation from the 
variation with respect to the lapse. The absence of this constraint seems 
to be the origin of a new ``mimetic'' degree of freedom. In this work, we
will not pursue this particular case further, leaving its 
interesting implications for future work.

In the next subsection~\ref{subsec:BG} we shall see how the physics can be
 changed/unchanged under a disformal transformation
 focusing first on the homogeneous and isotropic background.
And then the discussion is extended to deal with the general situation
 in the next subsection~\ref{subsec:NL} obtaining the same conclusion.

\subsection{Cosmological background \label{subsec:BG}}

Before going in depth let us clarify our starting point in order to 
avoid any confusion. We begin with a given action for a given metric and
 we are interested in the change of physics under a transformation of the metric. 
Obviously, we do not want to alter our model, i.e. our action. 
For this reason, we must take a passive approach. What we mean by a passive 
transformation is the following. For simplicity, let us consider two 
metrics $\bar{g}=d\bar{s}^2$ and $g=ds^2$ which are related by
a conformal transformation $\bar{g}=\Omega^2g$. 
Let us assume that we are given a model with the metric $\bar{g}$ 
and the action $S[\bar{g}]$. Now, we perform the transformation to the metric. 
Basically, we have two options. We can rewrite the action in terms of the 
transformed metric $S[\bar{g}]=S[\Omega^{2}g]$, or replace the metric 
$\bar{g}\rightarrow g=\Omega^{-2}\bar{g}$ while keeping the same functional 
form of the action, which yields a different value of the 
action $S[\bar{g}]\neq {S}[g]$. 
We call the former a passive transformation and the latter an active one. 
As we stated above, we consider the former. 
From this point of view, we should expect at most a change in the interpretation
 of the physics but not a change in observational results.
We will come back to this point later with an explicit form of the action 
in section \ref{sec:action}. 

That being said, we can readily have an idea of 
the effect of a disformal transformation by expressing our original line element
 in terms of the disformally transformed one.
 To be specific, we consider a model with a metric $\bar{g}$ 
and by means of a disformal transformation we choose to work in terms of $g$, 
i.e. we passively transform the barred frame to the unbarred frame by 
$\bar{g}=\bar{g}(g)$. 

Let us focus on a spatially homogeneous and isotropic
background with the metric,
\begin{align}
ds^2=g_{\mu\nu} (t) dx^\mu dx^\nu = -dt^2+g_{ij}dx^idx^j,
\label{eq:metric1}
\end{align}
where we have chosen the cosmic proper time, $g_{tt}=-1$ ($N=1$), and 
$g_{ij}=a^2(t) \Omega_{ij}$
 with $\Omega_{ij}$ 
being the metric of a homogeneous and isotropic 3-space.
The disformal transformation \eqref{eq:dis}\,,
\begin{align}
d\bar{s}^2&=\bar{g}_{\mu\nu} (t) dx^\mu dx^\nu
= \Bigl( g_{\mu\nu}+F(\phi,X)\phi_{,\mu}\phi_{,\nu} \Bigr) dx^\mu dx^\nu \,,
\end{align}
 with $N=1$ \eqref{eq:disdown} can be read as
\begin{align}
d\bar{s}^2
&=ds^2+ \Bigl( 1-\alpha^2(t) \Bigr) dt^2
 =-\alpha^2 (t)dt^2+g_{ij}dx^idx^j.
\label{eq:metric2}
\end{align}
It should be noted that $\alpha$ here depends only on the time coordinate.
This enables us to perform a time coordinate transformation given by
\begin{align}
d\bar{t}=\alpha(t) dt\,.
\label{eq:suit1}
\end{align}
With this new time coordinate, the barred metric is expressed 
simply as
\begin{align}
d\bar{s}^2 = -d\bar{t}^2+g_{ij}dx^idx^j\,.
\label{eq:metric3}
\end{align}
As clear from this, the time coordinate $\bar{t}$ is in fact 
the cosmic proper time in the barred frame.

One readily see that the above form of the metric is exactly in the
same form as the unbarred metric with the replacement $t\to\bar{t}$,
with the understanding that the scale factor in the transformed frame,
say $A(\bar{t})$, is regarded as a function of $\bar{t}$ through its 
$t$-dependence, $A(\bar{t})=a\bigl(t(\bar{t})\bigr)$. 
In the sense of the foregoing discussion, if one were to start from 
an action with the barred metric $\bar{g}_{\mu\nu}$, one would find that 
the form of the action in terms of the unbarred metric $g_{\mu\nu}$ can be 
interpreted as a rescaling of time from $t$ to $\bar{t}$, or in
a symbolic form $S[\bar{g}(g);t]=S[g;\bar{t}]$. See section \ref{sec:action} 
for an explicit proof. 

This indicates an important fact that a (cosmological) disformal 
transformation \eqref{eq:disdown} is essentially equivalent to a rescaling
of the time coordinate \eqref{eq:suit1}.
Since this rescaling applies to all the physical quantities equally
in the transformed frame, it implies the invariance of the physics under 
the disformal transformation. In other words, since we work with the same 
action but expressed in terms of new variables, observed physics should
not change.
If this reasoning is valid, it is straightforward that in particular
neither the causal structure nor the propagation speed are modified under
the disformal transformation. 

\paragraph{Wave propagation.}

We now consider the effect of a disformal transformation on
the propagation of waves. For simplicity, let us focus on a
scalar wave of comoving wavenumber $k$ living in the barred frame,
\begin{align}
\left[\frac{1}{a^3}\frac{d}{\bar{N}dt}\left(a^3\frac{d}{\bar{N}dt}\right)
+\frac{\bar{c}_s^2k^2}{a^2}\right]\bar{\phi}_k(t)=0\,,
\end{align}
where $\bar{c}_s$ is the sound velocity. Applying the passive disformal 
transformation \eqref{eq:disdown}, it becomes
\begin{align}
\left[\frac{1}{a^3\alpha}\frac{d}{Ndt}\left(\frac{a^3}{\alpha}\frac{d}{Ndt}\right)
+\frac{\bar{c}_s^2k^2}{a^2}\right]\bar{\phi}_k(t)
=
\frac{1}{\alpha^2}
\left[\frac{\alpha}{a^3}\frac{d}{dt}\left(\frac{a^3}{\alpha}\frac{d}{dt}\right)
+\frac{\alpha^2\bar{c}_s^2k^2}{a^2}\right]\bar{\phi}_k(t)=0\,,
\end{align}
where in the last step we set $N=1$ so that we chose $t$ to be 
the cosmic proper time of the unbarred frame.
Thus it appears that the sound velocity is changed to $c_{s,ap}=\alpha \bar{c}_s$, where $c_{s,ap}$ stands for apparent sound speed.
However, recalling that the cosmic proper time in the barred frame
is given by $d\bar{t}=\alpha dt$, the actual sound velocity
in the transformed frame should be read off from the equation
rewritten in terms of $\bar{t}$,
\begin{align}
\left[\frac{1}{a^3}\frac{d}{d\bar{t}}\left(a^3\frac{d}{d\bar{t}}\right)
+\frac{\bar{c}_s^2k^2}{a^2}\right]{\phi}_k(\bar{t})=0\,,
\end{align}
where $\phi_k(\bar{t})=\bar{\phi}_k\bigl(t(\bar{t})\bigr)$
or equivalently $\bar{\phi}_k(t)=\phi_k(\bar{t}(t))$. Here
it is important to note that no scalar 
function is modified by such a passive disformal transformation. 
Essentially, the functional form has apparently changed but not 
its value as indicated above.
It is apparent that the physical sound velocity $\bar{c}_s$ is the same in 
both frames. In fact, this is similar to that pointed out 
by Ellis and Uzan in \citep{ellis2005c,ellis2007note}. We further use this result 
in section \ref{sec:implications}. 

\paragraph{Causal structure.}
Bearing in mind the above result, let us discuss the causal structure. 
Consider the norm of a vector $k^\mu$ in the barred frame, i.e.
\begin{align}
k^2\equiv\bar{g}_{\mu\nu}\bar{k}^\mu\bar{k}^\nu
=-\bar{N}^{2}(\bar{k}^t)^2+g_{ij}\bar{k}^i\bar{k}^j,
\end{align}
where $k^2>0$, $=0$ or $<0$ for a spacelike, null or time-like vector,
respectively. Actually, $k^2$ is a scalar and consequently is invariant under
passive disformal transformations. By imposing such a condition, the invariance
of the causal structure is automatic. Conversely, the vector components must 
change in order to balance the transformation of the metric and to keep $k^2$
 invariant. 

Let us briefly discuss its consequences. The passive disformal 
transformation \eqref{eq:disdown} in $k^2$ leads us to
\begin{align}
k^2&=\bar{g}_{\mu\nu}\bar{k}^\mu\bar{k}^\nu
\cr
&=\left({g}_{\mu\nu}
+F(\phi,X)\phi_{,\mu}\phi_{,\nu}\right)\bar{k}^\mu\bar{k}^\nu
=-\alpha^{2}(\bar{k}^t)^2+g_{ij}\bar{k}^i\bar{k}^j\,,
\label{eq:k2trans}
\end{align}
where again we have set $N=1$ so that $t$ is the proper time in the
unbarred frame.
From the above we can identify the transformation rule for the vector 
components under the disformal transformation, that is
\begin{align}
\bar{k}^t=\alpha^{-1}{k}^t \quad {\rm and} \quad \bar{k}^i=k^i.
\label{eq:vectortransfup}
\end{align}
The covariant components transform according to
\begin{align}
\begin{array}{ccc}
\bar{k}_t=\alpha{k}_t &\quad {\rm and} &\quad \bar{k}_i=k_i.
\label{eq:vectortransf}
\end{array}
\end{align}
A straightforward consequence is that the sound speed is seemingly modified,
 i.e. 
\begin{align}
{c}^{-1}_s\equiv\frac{d{k}^t}{d|k|}=\alpha\frac{d\bar{k}^{t}}{d|k|}
=\alpha {c}^{-1}_{s,ap}~,
\end{align}
where ${c}_{s,ap}$ is the apparent sound speed, similar to the case of a scalar wave equation. 
Actually, this change in the sound speed is an artifact of working with 
time $t$ not proper to the barred frame. With the proper time $\bar{t}$
in the barred frame given by \eqref{eq:suit1}, we can rewrite $k^2$ as
\begin{align}
k^2=-(\bar{k}^{\bar{t}})^2+g_{ij}\bar{k}^i\bar{k}^j\,,
\end{align}
where $\bar{k}^{\bar{t}}=\alpha\bar{k}^t$. Notice that the physical sound speed is frame independent as well, namely
\begin{align}
\bar{c}^{-1}_s\equiv\frac{d\bar{k}^{\bar{t}}}{d|k|}=c_s^{-1}~.
\end{align}
 Before ending this subsection, 
let us stress again that we consider passive disformal transformations
throughout this paper unless otherwise stated.

\subsection{Non-linear considerations \label{subsec:NL}}

In this subsection we shall extend the previous discussion by taking
into account the spatial dependence of the scalar field,
 which leads to the spatial dependence of the metric and the disformal factor.
Here we assume that the perturbation expansion is valid, i.e., 
we assume the existence of a spatially homogeneous background solution
on which the physical spacetime can be constructed.
Note that, however, the perturbation can be fully nonlinear.
In this case, it is crucial to require the existence of a comoving slicing, 
i.e. the normal vector of constant $\phi$ hyper-surfaces is assumed to be
time-like everywhere, 
which ensures that time is the only component stretched by a disformal
 transformation. Even so, there is still a spatial dependence of the disformal 
factor through $X$ due to the fluctuations of the metric. 

Under this assumption, the previous discussion can be generalized, in practice,
 by introducing a local Lorentz frame where the spatial dependence of the 
disformal factor can be neglected since the scalar field can be considered 
spatially homogeneous in this frame. Consequently, the essential effect due 
to a disformal transformation is a mere change of the local Lorentz lapse 
function, which in turn can be absorbed by a redefinition of the local time 
coordinate. Hence, it implies the equivalence between a disformal transformation 
and a rescaling of time. In this way, the previous discussion for
the background can be generalised a priori to include non-linear perturbations
without any difficulty.

\section{Action invariance \label{sec:action}}

Thus far, we have shown that at the metric level a disformal transformation
 is equivalent to rescaling time. The next point to deal 
with is to show whether this expectation holds at the action level for 
gravitational and matter fields. Before going into details, let us clarify 
the notation. We start from a metric $\bar{g}_{\mu\nu}$, which we refer to 
it as the barred frame, and we consider the change of the action under a 
passive disformal transformation, that is we re-express the action in terms of 
the disformally related metric ${g}_{\mu\nu}$, which we call the unbarred frame.
 We shall do the same procedure for all fields.
In the next subsection we consider the Einstein-scalar action in 
the (3+1)-decomposition including nonlinear cosmological perturbations. 
Afterwards, we consider the action for several kinds of matter fields 
in the same decomposition of the metric.

\subsection{Einstein-scalar action \label{sec:grav}}
Let us address the disformal transformation of the gravitational action,
for the sake of simplicity the Einstein-Hilbert action with a canonical 
scalar field. At the end of this section we clearly show that the same 
procedure applies to its scalar-tensor theory extensions. For this purpose,
a crucial point is to work in the \textit{comoving}, or the
 \textit{uniform $\phi$ slicing} ($\delta\phi=0$)
 which ensures that only the time derivative of the scalar field plays
 a role in the disformal transformation as no perturbations of the scalar
 field are present on this slicing.\footnote{This is always possible in
 the single field case.} 
It is appropriate to work within the (3+1)-decomposition of the 
unbarred/barred metrics which are given by
\begin{align}
ds^2=-{N}^2(t, {\bm x})dt^2+ h_{ij} (t, {\bm x}) dx^idx^j
\quad \ma{and} \quad
 d\bar{s}^2=-\bar{N}^2(t, {\bm x})dt^2+ h_{ij} (t, {\bm x}) dx^idx^j,
\label{eq:adm}
\end{align}
where $N$, $\bar{N}$ and $h_{ij}$ are the unbarred/barred lapse and
the spatial metric, respectively.\footnote{We are using the fact that 
a disformal transformation does not affect the spatial components. 
We use $h_{ij}$ not to overload notation.} Note that we set to zero 
the shift vector, $\bar{N}^i=N^i=0$, by choosing a particular set of 
spatial coordinates. This fact will make the disformal transformation 
much clearer afterwards, though one can include a non-vanishing shift 
vector without essential difficulties.
 In this decomposition the action reads
\begin{align}
S_g=\frac{1}{2}\int d^3x dt \bar{N}\sqrt{h}\left\{R^{(3)}
+\bar{K}_{ij}\bar{K}^{ij}-\bar{K}^2
+\bar{N}^{-2}\left(\partial_{t}\phi\right)^2-2V(\phi)
+2\bar{\nabla}_\mu\left(\bar{n}^\mu\bar{\nabla}_\nu \bar{n}^\nu
-\bar{n}^\nu\bar{\nabla}_\nu \bar{n}^\mu\right)\right\},
\label{eq:gravaction}
\end{align}
where $^{(3)}R$, $\bar{K}_{ij}$ and $\bar{n}_\mu dx^\mu=-\bar{N}dt$ 
respectively are the spatial Ricci scalar, the extrinsic curvature and 
the normal vector of the spatial hyper-surface. The extrinsic curvature and 
the gradient of the normal vector given in terms of \eqref{eq:adm} read
\begin{align}
\begin{array}{cc}
K_{ij}=\frac{1}{2\bar{N}}\partial_t h_{ij}, &\quad \bar{K}=h^{ij}\bar{K}_{ij},
\end{array}
\end{align}
and
\begin{align}
\bar{\nabla}_\mu \bar{n}_\nu=\delta_\mu^0\delta_\nu^i \bar{N}_{,i}
+\delta^i_\mu\delta^j_\nu \bar{K}_{ij}.
\label{eq:gradn}
\end{align}
In this way, we can explicitly express the action in terms of 
the $\bar{N}$ and $h_{ij}$, i.e.
\begin{align}
S_g=\frac{1}{2}\int d^3x dt \bar{N}\sqrt{h}
\bigg\{R^{(3)}
&+\bar{K}_{ij}\bar{K}^{ij}-\bar{K}^2
+\bar{N}^{-2}\left(\partial_{t}\phi\right)^2-2V(\phi)
\cr
+&\frac{2}{\sqrt{h}\bar{N}}\partial_{t}(\sqrt{h}\bar{K})
 -\frac{2}{\sqrt{h}\bar{N}}\partial_{i}(\sqrt{h}h^{ij}\partial_j\bar{N})\bigg\},
\label{eq:barframe}
\end{align}
where we have kept the last two total derivative terms as they contribute 
if a non-minimal coupling is present. 

Let us now perform a disformal transformation 
to the unbarred frame where the action takes the form,
\begin{align}
\begin{split}
S_g=\frac{1}{2}\int d^3x dt \alpha N\sqrt{h}&\bigg\{R^{(3)}
+\alpha^{-2}\left({K}_{ij}{K}^{ij}-{K}^2
+{N}^{-2}\left(\partial_{t}\phi\right)^2\right)-2V(\phi)\\
&+\frac{2}{\sqrt{h}\alpha{N}}\partial_{t}(\sqrt{h}\alpha^{-1}{K})
-\frac{2}{\sqrt{h}\alpha{N}}
\partial_{i}\left(\sqrt{h}h^{ij}\partial_j(\alpha{N})\right)\bigg\},
\label{eq:gravaction2}
\end{split}
\end{align}
where we used the fact that $\bar{N}(t,x^i)=\alpha(\phi,X) N(t,x^i)$ 
from Eq.~\eqref{eq:disdown} and that ${K}_{ij}=\frac{1}{2{N}}\partial_t h_{ij}$. 
At this point, we have to be careful due to the hidden dependence of the 
old lapse in the disformal factor, i.e.
\begin{align}
\alpha(\phi,X)=\alpha(t,N)=\alpha(t,x^i),
\end{align}
which does not allow us to do a straightforward time redefinition. 
However, we can split the lapse and disformal factor into the background and 
the perturbed values. In our notation they are defined by
\begin{align}
\begin{array}{ccc}
N(t,x^i)={\rm e}^{n(t,x^i)}&\quad {\rm and} &\quad \alpha(t,x^i)
=\alpha_0(t){\rm e}^{\sigma(t,x^i)},
\label{eq:npert}
\end{array}
\end{align}
where $\alpha_0(t)$ is the background value of the disformal factor which 
is only a function of time. Likewise, the barred lapse is decomposed by 
$\bar{N}(t,x^i)=\alpha_0(t){\rm e}^{\bar{n}(t,x^i)}$. Note that thanks 
to this decomposition we can absorb the background disformal transformation
in a time redefinition \eqref{eq:suit1} given by 
\begin{align}
d\bar{t}=\alpha_0(t)dt.
\label{suit2}
\end{align}
As a result, physics is invariant \textit{at the background level}. 
It should be noted that this is valid as long as perturbation theory applies.

\paragraph{Implications for cosmological perturbations.} For simplicity 
we do not consider a non-minimal coupling here and, therefore, 
we drop the total derivative terms in \eqref{eq:gravaction2}. In this way,
the action in the unbarred frame is now given by
\begin{align}
\begin{split}
S_g=\frac{1}{2}\int d^3x\,d\bar{t}\sqrt{h}\,{\rm e}^{{n}(t,x^i)+\sigma(t,x^i)}
\bigg\{R^{(3)}
+{\rm e}^{-2 [{n}(t,x^i)+\sigma(t,x^i)]}\Bigl(E_{ij}E^{ij}-E^2
+(\partial_{\bar{t}}\phi)^2\Bigr)-2V(\phi)\bigg\},
\label{eq:gravdisf}
\end{split}
\end{align}
where we defined $E_{ij}\equiv\frac{1}{2}\partial_{\bar{t}}h_{ij}$ 
and $E=h^{ij}E_{ij}$, following the notation of Maldacena \citep{maldacena2003non},
so that every quantity is expressed in terms of the new time $\bar{t}$ but 
for the perturbed lapse ${n}(t,x^i)$ and the perturbed disformal factor 
$\sigma(t,x^i)$. We deliberately kept the old time dependence in the latter
for the reason shown below. In the Hamiltonian formalism, the lapse
is a Lagrange multiplier and, hence, a redefinition of it has no effect
in the dynamics. For this reason, let us redefine the lapse so that it 
absorbs the perturbed disformal factor, i.e.
\begin{align}
{n}(t,x^i)+\sigma(t,x^i)=\bar{n}(t,x^i).
\label{eq:barn}
\end{align}
In terms of this lapse, the Hamiltonian constraint is given by
\begin{align}
R^{(3)}
-{\rm e}^{-2\bar{n}(t,x^i)}\Bigl[E_{ij}E^{ij}-E^2+(\partial_{\bar{t}}\phi)^2
 \Bigr]-2V(\phi)=0,
\end{align}
which at the background level has the same solution as \eqref{eq:gravaction} 
but given in terms of the redefined time $\bar{t}$. Due to this fact, 
the perturbed barred lapse must be exactly equal to the unbarred one but 
expressed in terms of the redefined time $\bar{t}$, in other words
\begin{align}
\bar{n}(t,x^i)=n(\bar{t},x^i).
\end{align}
Consequently, the action \eqref{eq:gravdisf} takes exactly the same form 
as \eqref{eq:gravaction} but in terms of the redefined time $\bar{t}$. 

We may go further and consider a non-minimal coupling or a Horndeski-type
Lagrangian. The lapse ${N}$ is always accompanied by a factor 
$\alpha$ which at the background level is successfully absorbed by the 
time redefinition and at the perturbation level gives rise to a factor 
${\rm e}^{\bar{n}(t,x^i)}$. 
For example, consider one of the total derivative terms in \eqref{eq:gravaction2} 
which in presence of a non-minimal coupling gives a non-trivial
contribution to the action as
\begin{align}
S_g \supset\int d^3x dt \, \alpha{N} \sqrt{h} \, \Omega(\phi,X)
\frac{1}{\sqrt{h}\alpha{N}}\partial_{t}(\sqrt{h}\alpha^{-1}{K}),
\end{align}
where $\Omega(\phi,X)$ is the non-minimal coupling. Absorbing the background 
value of the disformal function and integrating by parts we are led to
\begin{align}
S_g \supset -\int d^3x d\bar{t}
\sqrt{h}\,{\rm e}^{-\bar{n}(t,x^i)} E~\partial_{\bar{t}}\Omega,
\end{align}
where the factor $X$ changes as well according to
\begin{align}
X=\frac{1}{2\alpha^2{N}^2}(\partial_t\phi)^2
=\frac{{\rm e}^{-2\bar{n}(t,x^i)}}{2}(\partial_{\bar{t}}\phi)^2.
\end{align}
In this way, the function $\bar{n}(t,x^i)$ appears exactly in the same 
place where the original lapse $n(t,x^i)$ is in the original 
action \eqref{eq:gravaction}. The only difference is the time coordinate 
$\bar{t}$ which is used in any other variables except for the perturbed lapse, which gives
us no other choice but to conclude that $\bar{n}(t,x^i)=n(\bar{t},x^i)$. \\

In summary, a cosmological disformal transformation in 
the \textit{comoving}, or the \textit{uniform $\phi$ slicing}
 is equivalent to a rescaling of the time 
coordinate even at \textit{higher orders} in perturbative expansion. 
Therefore, as long as perturbation theory is valid, cosmological 
perturbations are invariant under a disformal transformation.
 This result is in agreement 
with \citep{creminelli2014resilience,minamitsuji2014disformal,tsujikawa2014disformal}
where it was found that scalar and tensor power spectra are frame 
independent at leading order in the perturbation.

\subsection{Towards a Horndeski Lagrangian\label{sec:horn}}
Once we know how the gravitational action transforms, it is 
appropriate to take a look at the resulting Horndeski-type action if 
we stick to the original time, i.e., without rescaling the time coordinate.
The beyond-Horndeski or GLPV Lagrangian can be written in terms of geometrical 
quantities \cite{gleyzes2013essential,gleyzes2014healthy,gleyzes2014exploring} 
and it consists of the following four terms:
\begin{flalign}
&{\cal L}_2=A_2(\phi,X), \\
&{\cal L}_3=A_3(\phi,X)K,\\
&{\cal L}_4=A_4(\phi,X)R^{(3)}+B_4(\phi,X)(K_{ij}K^{ij}-K^2), \\
&{\cal L}_5=A_5(\phi,X)\left(R^{(3)}_{ij}K^{ij}-\frac{R^{(3)}K}{2}\right)
+B_5(\phi,X)\left(K^3-3K K_{ij}K^{ij}+2K_{ij}K^{ik}K^j_k\right).
\end{flalign}
The GLPV Lagrangian reduces to the Horndeski Lagrangian when the 
functions $B_4$ and $B_5$ are respectively related to $A_4$ and $A_5$ by
\begin{align}
\begin{array}{ccc}
B_4(\phi,X)=A_4-2XA_{4,X} &\quad {\rm and}&\quad B_5(\phi,X)
=-\dfrac{1}{2}XA_{5,X}.
\label{eq:horndeskicond}
\end{array}
\end{align}
A glance at the disformally transformed action in terms of the original 
time \eqref{eq:gravaction2} tells us that the existence of an Einstein 
frame from the Horndeski theory point of view
 is closely related to the absence of the term ${\cal L}_5$.
 In addition, we can identify the following terms for 
a cosmological background with a canonical scalar field with a potential:
\begin{align}
A_2(\phi,X) =\alpha^{-1}X-\alpha V\,, \quad A_3(\phi,X)=0 \,, \quad
 A_4(\phi,X)=\alpha \quad {\rm and} \quad B_4(\phi,X)=\alpha^{-1},
\label{eq:hornrelations}
\end{align}
where we remind the reader that $\alpha(\phi,X)=\sqrt{1-2XF(\phi,X)}$.
 The first condition in \eqref{eq:horndeskicond}, which a Horndeski 
Lagrangian must satisfy, implies that the disformal factor must fulfil
\begin{align}
\alpha^{-1}=\alpha-2X\alpha_{,X},
\label{eq:hornrelations2}
\end{align}
which means $F(\phi,X)=F(\phi)$. As a result, only
 a \textit{field dependent} disformal transformation leads us to a Horndeski 
Lagrangian. This is also pointed out by Bettoni and Liberati 
in \citep{bettoni2013disformal} with a general treatment. 
In other words, a general disformal transformation of the Einstein-scalar
theory does not satisfy the condition that the field equations are
at most second order in time derivatives. This implies the existence
of higher time derivatives and hence of Ostrogradsky ghosts.
Nevertheless, despite the apparent appearance of Ostrogradsky ghosts for
 a $X$ dependent disformal transformation, no real ghost should actually be 
present as the theory in the transformed frame is equivalent to
the healthy Einstein-scalar theory. This is discussed by 
Zumalac\' arregui and Garc\' ia-Bellido 
in \citep{ zumalacarregui2014transforming} where they show the existence
 of hidden constraints. We leave its application to cosmology for future work.

\subsection{Matter action\label{sec:matter}}
Now let us consider matter fields.
 We work under the same decomposition of the metric given by 
 \eqref{eq:adm} but only considering the background dynamics. 
As explained in section \ref{subsec:NL}, the generalisation to include 
perturbations is straightforward.

\paragraph{Scalar field.}
Let us start with a scalar field $\chi$ with mass $m$ whose action, 
in the barred frame, is given by 
\begin{align}
S=-\frac{1}{2}\int d^4x \sqrt{-\bar{g}}
\left(\bar{g}^{\mu\nu}
\partial_\mu\chi\partial_\nu\chi+m^2\chi^2\right),
\label{eq:scalaraction}
\end{align}
which in terms of the barred metric \eqref{eq:adm} yields
\begin{align}
S=\frac{1}{2}\int d^3x~dt \bar{N}\sqrt{\bar{h}}
\, \Bigl\{ \bar{N}^{-2}
(\partial_t\chi)^2-\bar{h}^{ij}
\partial_i\chi\partial_j\chi-m^2\chi^2 \Bigr\}.
\label{eq:scalaraction1}
\end{align}
By means of a disformal transformation \eqref{eq:disdown} we find that 
the action in the unbarred frame reads
\begin{align}
S=\frac{1}{2}\int d^3x~ dt~\alpha N \sqrt{h} 
\, \Bigl\{ N^{-2}\alpha^{-2}
(\partial_t\chi)^2-h^{ij}
\partial_i\chi\partial_j\chi-m^2\chi^2 \Bigr\},
\label{eq:scalaraction2}
\end{align}
where we used that $\bar{N}(t)=\alpha(t) N(t)$. Note that one may argue 
that the sound speed in the unbarred frame is modified compared to that in
 the barred frame. However, there is a factor $\alpha$ in front of every $dt$
 which can be successfully absorbed by the time redefinition \eqref{eq:suit1} 
that leads us to
\begin{align}
S=\frac{1}{2}\int d^3x\,d\bar{t}\,N\sqrt{h} 
\, \Bigl\{ N^{-2}(\partial_{\bar{t}}\chi)^2-h^{ij}
\partial_i\chi\partial_j\chi-m^2\chi^2 \Bigr\},
\label{eq:scalaraction3}
\end{align}
which is the action one would expect in the unbarred frame \eqref{eq:adm} 
with the relabelling $t\to\bar{t}$. In other words, if we
denote a solution of the field equation in the barred frame by $\bar\chi(t)$,
the corresponding solution in the unbarred frame $\chi(t)$ is related
to it as $\chi(\bar{t})=\bar\chi(t)$. 

\paragraph{Electromagnetic field.}
Let us move towards the disformal transformation of the electromagnetic field.
 It is known that the Maxwell field is conformally invariant \citep{wald2010general}
 but in general it may not be disformally invariant. This is an important point 
since observations mainly use photons. However, we want to stress that as 
long as matter is universally coupled to a unique metric, no issue arises (see appendix \ref{sec:observables} for an example). The action for the electromagnetic
 field $A_\mu$ in the barred frame reads
\begin{align}
S=-\frac{1}{4}\int d^4x 
\sqrt{-\bar{g}}\bar{g}^{\alpha\mu}\bar{g}^{\beta\nu}F_{\alpha\beta}F_{\mu\nu},
\label{eq:maxwellaction}
\end{align}
where $F_{\alpha\beta}=\partial_\alpha A_\beta-\partial_\beta A_\alpha$. 
We expand the action \eqref{eq:maxwellaction} in terms of the 
barred metric \eqref{eq:adm} to explicitly split the time components from the 
spatial ones, i.e.
\begin{align}
S=\frac{1}{4}\int d^3x \,dt\,\bar{N}\sqrt{\bar{h}}
\left(2\bar{N}^{-2}\bar{h}^{ij}F_{ti}F_{tj}
-\bar{h}^{ij}\bar{h}^{kl}F_{ik}F_{jl}\right),
\label{eq:maxwellaction1}
\end{align}
and perform the disformal transformation \eqref{eq:disdown} to obtain
\begin{align}
S=\frac{1}{4}\int d^3x\,dt\alpha N \sqrt{h}
\left(2\alpha^{-2}N^{-2}h^{ij}F_{ti}F_{tj}-h^{ij}h^{kl}F_{ik}F_{jl}\right).
\label{eq:maxwellaction2}
\end{align}
Note that $F_{ti}=\partial_t A_i-\partial_i A_t$ and, therefore, in order 
to successfully absorb the $\alpha$ factor in the time 
redefinition \eqref{eq:suit1} the time component of
the electromagnetic field must be transformed as
\begin{align}
A_{\bar{t}}=\alpha^{-1}A_t\,.
\label{eq:At}
\end{align}
The resulting action is given by
\begin{align}
S=\frac{1}{4}\int d^3x\,d\bar{t}\,N 
\sqrt{h}\left(2N^{-2}h^{ij}F_{\bar{t}i}F_{\bar{t}j}-h^{ij}h^{kl}F_{ik}F_{jl}\right),
\label{eq:maxwellaction3}
\end{align}
where we have used the fact that
 $\pa_i A_{\bar{t}} = \alpha^{-1}\pa_i A_t$ for a spatially
 homogeneous $\alpha=\alpha(t)$. Again this action is the one for the 
electromagnetic field in the metric \eqref{eq:adm} but labelled 
in terms of $\bar{t}$.
 
The redefinition of the electromagnetic field given by equation \eqref{eq:At}
 is not surprising within the differential form approach, 
where the 1-form is given by
\begin{align}
{\mathbb{A}}&=A_\mu dx^\mu=A_{t} dt+A_idx^i \notag\\
&=A_t \alpha^{-1} d\bar{t}+A_idx^i=A_{\bar{t}} d\bar{t}+A_idx^i.
\label{eq:oneform}
\end{align}
Thus, the 1-form $\mathbb{A}$ is invariant under cosmological disformal 
transformations.

\paragraph{Dirac field.}
For the sake of completeness let us quickly consider the disformal
transformation of a Dirac field, e.g. an electron. The action
for a fermion field $\Psi$ with mass $m$ and charge $e$ in curved
 space-time is given by \citep{birrell1984quantum}
\begin{align}
S=-\frac{i}{2}\int d^4x \sqrt{-\bar{g}}
\left(\breve{\Psi}\bar{\gamma}^{\mu}
\bar{D}_\mu\Psi+m\breve{\Psi}\Psi\right),
\label{eq:diracaction}
\end{align}
where $\breve{\Psi}=\Psi^\dagger\gamma^0$ is the adjoint spinor,
 $\bar{D}_\mu=\partial_\mu+ieA_\mu+\bar{\Gamma}_\mu$ and
 the tetrad components $\bar{e}^\mu_{(a)}$ are defined by
\begin{align}
\eta_{ab}=\bar{e}^\mu_{(a)}\bar{e}^\nu_{(b)}\bar{g}_{\mu\nu},
\label{eq:tetrad}
\end{align}
which relate the gamma matrices and the spin connection in curved space-time
to the gamma matrices in flat space-time, i.e.
 $\bar{\gamma}^\mu=\gamma^a \bar{e}^\mu_{(a)}$ and
 $\bar{\Gamma}_\mu=\frac{1}{8}[\gamma^a,\gamma^b]\bar{e}^\lambda_{(a)}
\bar{\nabla}_\mu\bar{e}_{(b)\lambda}$. In the (3+1)-decomposition the action
 reads
\begin{align}
S=-\frac{i}{2}\int d^3x~dt~\bar{N}\sqrt{\bar{h}}
\left(\breve{\Psi}\gamma^0 \bar{N}^{-1}
\bar{D}_t\Psi+\breve{\Psi}\bar{\gamma}^{i}
\bar{D}_i\Psi+m\breve{\Psi}\Psi\right),
\label{eq:diracaction1}
\end{align}
where $\bar{e}^t_{(a)}=\delta^0_{(a)}\bar{N}^{-1}$ and 
$\bar{e}^i_{(a)}\bar{e}^j_{(b)}=\delta_{ab}\bar{h}^{ij}$ 
so that $\bar{\gamma}^t=\gamma^0 \bar{N}^{-1}$.
The effect of the disformal transformation \eqref{eq:disdown} can be
summarised as follows. The tetrad is modified according to \eqref{eq:tetrad},
\begin{align}
\begin{array}{ccc}
\bar{e}^t_{(a)}=e^t_{(a)}\alpha^{-1} &\quad {\rm and}&\quad
 \bar{e}^i_{(a)}=e^i_{(a)},
\label{eq:distetrad}
\end{array}
\end{align}
which yields a similar transformation for the gamma matrices,
\begin{align}
\begin{array}{ccc}
\vspace{1mm}
\bar{\gamma}^t=\gamma^{t}\alpha^{-1}\,, &\quad &\quad
 \bar{\gamma}^i=\gamma^i.
\end{array}
\label{eq:diracdisformal}
\end{align}
The modification of the spin connection $\bar{\Gamma}_\mu$ is slightly 
more involved. Essentially, there is a contribution not only 
from the transformation of the tetrad $\bar{e}^\lambda_{(a)}$ but also
from its covariant derivative $\bar{\nabla}_\mu\bar{e}_{(b)\lambda}$. 
The latter is due to the change of the Christoffel symbols, which are defined by 
$\Gamma^{\lambda}_{\mu\nu}=\frac{1}{2}
\bar{g}^{\lambda\sigma}\left(\partial_\mu\bar{g}_{\sigma\nu}
+\partial_\nu\bar{g}_{\mu\sigma}-\partial_\sigma\bar{g}_{\mu\nu}\right)$.
 Specifically, the non-vanishing Christoffel symbols for the
 metric \eqref{eq:metric2} which are affected by the disformal 
transformation are explicitly given by
\begin{align}
\begin{array}{ccc}
\vspace{2mm}
\bar{\Gamma}^t_{tt}=\frac{1}{2}\bar{N}^{-2}\partial_t\bar{N}^2\,, &\quad &\quad
\bar{\Gamma}^i_{tt}=\frac{1}{2}\bar{h}^{ij}\partial_j \bar{N}^{2}~, \\
\vspace{2mm}
\bar{\Gamma}^t_{it}=\frac{1}{2}\bar{N}^{-2}\partial_i \bar{N}^{2}\,, &\quad &\quad
\bar{\Gamma}^t_{ij}=\frac{-1}{2}\bar{N}^{-2}\partial_t \bar{h}_{ij}~, \\
\vspace{0mm}
\bar{\Gamma}^i_{jt}=\frac{1}{2}\bar{h}^{ik}\partial_t \bar{h}_{kj}~,
\end{array}
\label{eq:christoffel}
\end{align}
where the spatial component $\bar{\Gamma}^i_{jk}$ is not modified under
 such a disformal transformation. This result greatly simplifies the expression 
for the spin connection. The time component $\bar{\Gamma}_t$ is proportional to
\begin{align}
\bar{e}^\lambda_{(a)}\bar{\nabla}_t\bar{e}_{(b)\lambda}
=\bar{e}^t_{(a)}\partial_t\bar{e}_{(b)t}
+\bar{e}^i_{(a)}\partial_t\bar{e}_{(b)i}
-\bar{e}^t_{(a)}\bar{\Gamma}^t_{tt}\bar{e}_{(b)t}
-\bar{e}^i_{(a)}\bar{\Gamma}^j_{ti}\bar{e}_{(b)j}
-\bar{e}^t_{(a)}\bar{\Gamma}^i_{tt}\bar{e}_{(b)i}
-\bar{e}^i_{(a)}\bar{\Gamma}^t_{ti}\bar{e}_{(b)t}\,,
\label{eq:spint}
\end{align}
and the spatial component $\bar{\Gamma}_j$ is proportional to
\begin{align}
\bar{e}^\lambda_{(a)}\bar{\nabla}_j\bar{e}_{(b)\lambda}
=\bar{e}^t_{(a)}\partial_j\bar{e}_{(b)t}+\bar{e}^i_{(a)}\partial_j\bar{e}_{(b)i}
-\bar{e}^t_{(a)}\bar{\Gamma}^t_{jt}\bar{e}_{(b)t}
-\bar{e}^i_{(a)}\bar{\Gamma}^k_{ji}\bar{e}_{(b)k}
-\bar{e}^i_{(a)}\bar{\Gamma}^t_{ij}\bar{e}_{(b)t}
-\bar{e}^t_{(a)}\bar{\Gamma}^i_{jt}\bar{e}_{(b)i}\,.
\label{eq:spini}
\end{align}

The dependence of $\Gamma_\mu$ on $\alpha$ after 
the disformal transformation \eqref{eq:disdown} can be
seen from the above two equations. 
Let us first look at the time component, i.e.
\begin{align}
\bar{\Gamma}_t=\frac{1}{8}[\gamma^a,\gamma^b]\,\bar{e}^\lambda_{(a)}
\bar{\nabla}_t\bar{e}_{(b)\lambda}.
\end{align}
From equation \eqref{eq:spint} we see that any term containing a time derivative,
namely the first four terms on the right hand side of it, has
no $\alpha$ dependence. In particular, the $\partial_t \alpha$ term arising 
from $\partial_t\bar{e}_{(b)t}$ is canceled with that coming from 
the $\bar{\Gamma}^t_{tt}$ term. The last two terms on the right hand side 
contain spatial derivatives and each spatial derivative
is accompanied by a factor $\alpha$. 
Next let us move to the spatial component, i.e.
\begin{align}
\bar{\Gamma}_j=\frac{1}{8}[\gamma^a,\gamma^b]\,
\bar{e}^\lambda_{(a)}\bar{\nabla}_j\bar{e}_{(b)\lambda}.
\end{align}
From \eqref{eq:spini} we realise that the first four terms on the right hand side
of it which contain spatial derivatives has no $\alpha$ dependence. 
On the other hand, those which contain time derivatives, i.e. the last two terms, 
have a factor $\alpha^{-1}$ for each time derivative. 
At the end of the day, we can schematically express the effect of
a disformal transformation on the spin connection as
\begin{align}
\begin{array}{ccc}
\vspace{1mm}
\bar{\Gamma}_t[\partial_t, \partial_i]=\Gamma_t[\partial_t,\alpha \partial_i]
=\alpha\Gamma_t[\alpha^{-1}\partial_t,\partial_i]
&\quad {\rm and}&\quad
\bar{\Gamma}_i[\partial_t,\partial_i]=\Gamma_i[\alpha^{-1}\partial_t,\partial_i],
\end{array}
\label{eq:diracdisformal2}
\end{align}
where 
$\Gamma_\mu=\frac{1}{8}[\gamma^a,\gamma^b]~e^\lambda_{(a)}\nabla_\mu e_{(b)\lambda}$
is the spin connection in the unbarred frame. 
It is clear now that the factor $\alpha^{-1}$ in $\bar{N}^{-1}$ in
front of $\bar{D}_t$ in \eqref{eq:diracaction1}
 cancels the factor $\alpha$ from $\bar{\Gamma}_t$,
and the factor $\alpha^{-1}$ associated with each $\partial_t$
is successfully absorbed in the time redefinition \eqref{eq:suit1}.
As a result, the disformal transformation of the full action can be rewritten as
 \begin{align}
S=-\frac{i}{2}\int d^3x\,d\bar{t}\,N \sqrt{h}
\left(\breve{\Psi}\gamma^0 N^{-1}
D_{\bar{t}}\Psi+\breve{\Psi}\gamma^{i}
D_i\Psi+m\breve{\Psi}\Psi\right),
\label{eq:diracaction2}
\end{align}
where $D_{\bar{t}}=\pa_{\bar{t}}+ieA_{\bar{t}}+\Gamma_{\bar{t}}$, $\Gamma_{\bar{t}}\equiv\Gamma_t[\partial_{\bar{t}},\partial_i]$
 and we 
have used the previous result for the electromagnetic field \eqref{eq:At}. 
In this form, the disformal invariance is certainly manifest.
Lastly, we note that neither the charge $e$ nor the mass $m$ 
is affected by a disformal transformation.

\section{Frame independence \label{sec:implications}}

In this section, we shall consider the system, gravity plus matter, as a whole. We assume that matter 
is \textit{universally coupled} to $\bar{g}_{\mu\nu}$, which we call the
matter frame, and is related by a disformal transformation to $g_{\mu\nu}$
which we call the gravitational frame. For example, we may have
the Einstein-Hilbert action for the metric $g_{\mu\nu}$.
Thus, the weak equivalence principle is preserved. This is in contrast 
with \citep{van2013constraints,brax2013cosmological} where different 
disformal couplings for matter and radiation are considered.

\subsection{Gravity plus matter \label{sec:application}}

In the preceding section we explicitly showed the invariance under
 disformal transformations field by field. In this subsection, we 
consider the whole system. For example, the simplest and commonly used 
action \citep{brax2013cosmological,van2013constraints,koivisto2012screening,
sakstein2014disformal,sakstein2015towards,moffat2003bimetric,magueijo2003new,
magueijo2009bimetric} is given by
\begin{align}
S=\int d^4x \, \Bigl\{ \sqrt{-g}\Bigl(R[g]+{\cal L}_\phi(g,\phi)\Bigr)
+ \sqrt{-\gamma} {\cal L}_m(\gamma,\psi_I) \Bigr\},
\label{eq:disaction}
\end{align}
where the gravitational sector is the Einstein-Hilbert action in terms of 
the metric $g$ plus and a scalar field $\phi$,\footnote{In principle, 
one could choose any form for the gravitational 
sector \citep{zumalacarregui2013dbi}.} the latter being responsible for
 a disformal coupling with matter fields $\psi_I$ through
\begin{align}
\gamma_{\mu\nu}=g_{\mu\nu}+H(\phi,X)\phi_{,\mu}\phi_{,\nu}\,.
\label{eq:dis1}
\end{align}
Under the cosmological assumption we adopt throughout this paper,
this amounts to the replacement of the lapse function in the matter sector by
\begin{align}
{N}_\gamma^2=N_g^2 \Bigl(1-2XH(\phi,X) \Bigr)\equiv N_g^2\beta^2(\phi,X)\,.
\label{eq:Nbar}
\end{align}
One may wonder whether the system is still disformally invariant as a whole.
 Let us show that this is the case. 

Let us perform a disformal transformation given by
\begin{align}
{g}_{\mu\nu}=\bar{g}_{\mu\nu}+F(\phi,\bar{X})\phi_{,\mu}\phi_{,\nu}\,,
\label{eq:dis2}
\end{align}
which introduces a new lapse function related to the original one by
\begin{align}
N_g^2=\bar{N}_g^2(1-2XF(\phi,\bar{X}))\equiv\bar{N}_g^2\alpha^2(\phi,\bar{X})\,,
\end{align}
where $\bar{X}=(\partial_t\phi)^2/{2\bar{N}_g^2}$.
As before, we introduce a new time coordinate by $d\bar{t}=\alpha_0(t)dt$
and absorb the perturbation of $\alpha$ into $\bar{N}_g$
by redefining it. This means
\begin{align}
N_gdt=\bar{N}_g \alpha dt = e^{\bar{n}_g} \alpha_0(t)dt
 = e^{\bar{n}_g} d\bar{t}\,,
\label{eq:Ngdt}
\end{align}
where we used equation \eqref{eq:npert} to define $e^{\bar{n}_g}\equiv\bar{N}_ge^{\sigma(t,x^i)}$. To apply the passive disformal transformation to the action 
we first express the matter sector in terms of $N_g$,
\begin{align}
S=\int d^3x~dt N_g \sqrt{h} \, \Bigl\{ R[g]+{\cal L}_\phi(g,\phi)
+\beta(\phi,X)
{\cal L}_m \bigl(h,(\beta N_g)^{-1}\pa_t \psi_I,\cdots \bigr)\Bigr\}\,.
\label{eq:disaction2}
\end{align}
Using (\ref{eq:Ngdt}), we can then rewrite the action in terms
of the barred time as
\begin{align}
S=\int d^3x~d\bar{t}\bar{N}_g \sqrt{h}
\, \Bigl\{ R[\bar{g}]+{\cal L}_\phi(\bar{g},\phi)
+\beta(\phi,X_{\bar{g}})
{\cal L}_m \bigl(h,(\beta\bar{N}_g)^{-1}\pa_{\bar{t}}\psi_I,\cdots \bigr)\Bigr\}\,,
\label{eq:disaction3}
\end{align}
where it should be noted that $X_{\bar{g}}$ is in fact 
equal to $X$ but rewritten in terms of barred quantities,
\begin{align}
X_{\bar{g}}\equiv\frac{2}{\bar{N}_g^2}
\left(\frac{\partial\phi}{\partial\bar{t}}\right)^2=X\,.
\end{align}
Once the disformal transformation is done we can go back to the original 
form of the action but in terms of barred quantities, i.e.
\begin{align}
S=\int d^3x~d\bar{t} \, \Bigl\{\sqrt{-\bar{g}}
 \Bigl(R[\bar{g}]+{\cal L}_\phi(\bar{g},\phi)\Bigr)
+\sqrt{-\bar{\gamma}}{\cal L}_m(\bar{\gamma},\psi_I)\Bigr\}\,,
\label{eq:disaction4}
\end{align}
by defining $\bar{\gamma}$ through 
$\bar{N}_\gamma\equiv\bar{N}_g\beta(\phi,X_{\bar{g}})$ or alternatively
\begin{align}
\bar{\gamma}_{\mu\nu}=\bar{g}_{\mu\nu}+H(\phi,X_{\bar{g}})\phi_{,\mu}\phi_{,\nu}~,
\label{eq:dis3}
\end{align}
which is identical to the original relation \eqref{eq:dis1} but 
in terms of barred quantities.

\subsection{Propagation speed of gravitons and photons{\label{gravitons}}}
One effect that we may be able to detect which is not present 
in the case of a conformal coupling was already pointed out by
 Bekenstein \citep{bekenstein1993relation}. It was argued that gravitons 
may travel faster or slower than light, depending on the sign of the 
disformal factor \eqref{eq:dis}. This is true when matter and gravity
are expressed in terms of two different metrics which are related
by a disformal transformation like \eqref{eq:dis1}. 
However, as we have seen, we should keep in mind that 
the disformal transformation itself does not change the propagation speed.
The difference in the propagation speed exists independent of frames.

To illustrate this fact let us consider the previous action given by
\begin{align}
S=\int d^4x \, \Bigl\{\sqrt{-g}\Bigl(R[g]+{\cal L}_\phi(g,\phi)\Bigr)
+ \sqrt{-\bar{g}} {\cal L}_m(\bar{g},\psi_I)\Bigr\},
\end{align}
where the propagation speed of gravitons may be different 
from that of photons, for example, due to the disformal
coupling. Note that we renamed $\gamma$ as $\bar{g}$ in order to 
be consistent with notation in section \ref{sec:action}.
Let us express the gravitational sector 
in terms of the matter metric as
\begin{align}
S=\int d^4x \, \Bigl\{\sqrt{-g[\bar{g}]}
 \Bigl(R[g[\bar{g}]]+{\cal L}_\phi(g[\bar{g}],\phi)\Bigr)
+ \sqrt{-\bar{g}} {\cal L}_m(\bar{g},\psi_I)\Bigr\}\,,
\end{align}
as given by the disformal transformation from (\ref{eq:barframe}) 
to (\ref{eq:gravaction2}), interchanging the roles of
the barred and unbarred metrics (see appendix \ref{sec:appendix}).
This gives rise to a Horndeski action schematically given by
\begin{align}
S=\int d^4x \sqrt{-\bar{g}} \, \Bigl\{{\cal L}_{\small horn}
(\bar{g},\bar{K},\bar{R},\phi,\bar{X})+ {\cal L}_m(\bar{g},\psi_I)\Bigr\}.
\end{align}
The interpretation in this form is different. Now, the relative difference
 of the propagation speed between gravitons and photons is not due to 
the disformal coupling of matter but due to a modification of gravity. 
The resulting Horndeski action in terms of the matter metric has
 a tensor propagation speed given by $c_T=\alpha^{-1}$.\footnote{Alternatively,
 it can be inferred from the relative factor $\alpha^2$ relating both 
metrics \eqref{eq:disdown}.} 
However, we emphasize that this is not an observable feature
 in the primordial power spectrum \citep{creminelli2014resilience},
which may be understood by the fact that the coupling in
the matter sector is irrelevant during inflation. 

The above discussion has an important implication.
It means that as long as the tensor power spectrum is concerned,
the tensor propagation speed may be always set to unity by 
a disformal transformation, a definition for the Einstein 
frame (at least perturbatively). This consequence has been also discussed
by Creminelli et al. \cite{creminelli2014resilience}. 
In any case, a disformal coupling with matter should be in principle 
observable if we could measure the relative difference in the
propagation speed between gravitons and matter, which is frame 
independent.

In this sense, if one is interested in the causal structure of
the whole system, by means of a disformal transformation one may
go to the frame where the propagation speed of the fastest species 
(e.g. either gravitons or photons) is equal to unity,
 i.e. equal to the geometrical factor $c$ which defines the causal structure.
In this frame, any other field will have a propagation speed less than 
unity and hence the standard causal structure is preserved.

\section{Conclusions\label{sec:conclusions}}

Disformal transformations play an important role as a generalization of 
conformal transformations. The latter is widely used in modifications of
 gravity as well as in cosmology and it has been proven to leave observable
physics invariant. In this paper, we focused on pure disformal transformations,
i.e. those without any transformation in the conformal factor.
Further we assumed a cosmological setting, i.e. under the assumption that 
a scalar field responsible for the disformal transformation is regular 
everywhere and its derivative is time-like,
which allows us to choose the time slicing on which the scalar field is 
 homogeneous. Any posterior conformal transformation can be independently 
done in general. Our main result is that such a disformal transformation,
which we call a cosmological disformal transformation, 
is equivalent to a rescaling of the time coordinate, which is valid
full non-linearly, provided that perturbation theory is applicable. 
From this result, it follows that observable physics must be invariant 
under cosmological disformal transformations. 
In this sense, we extended the work for cosmological perturbations in the 
literature \citep{minamitsuji2014disformal,tsujikawa2014disformal} which 
showed the disformal invariance at leading order in the perturbation.
We placed emphasis on the fact that neither the causal structure, propagation 
speed nor any other property of the fields is affected by a disformal 
transformation itself, although it may seem so if an appropriate
time is not used.

The physical invariance under disformal transformations is an interesting 
result. It may help us to better understand Horndeski or Galileon theory.
It may simplify calculations by going to the frame where the speed of sound 
of the tensor modes is equal to unity, i.e. to the Einstein frame. 
For example, if applied to inflation, the subset of Horndeski theory with coefficients \eqref{eq:hornrelations} satisfying \eqref{eq:hornrelations2} is observationally indistinguishable from the corresponding Einstein-Hilbert action with a canonical scalar field with a potential.

On the other hand, if applied to dark energy where gravity and matter
may have different disformal couplings, there are strong constraints 
from the solar system experiments due to the appearance of a fifth 
force \citep{sakstein2014disformal}. Nevertheless, a slowly rolling field 
may be able to pass those constraints as the disformal transformation 
depends on its first time derivative \citep{sakstein2015towards}. 
Another observationally distinguishable signature is the relative difference 
of the propagation speed between gravitons and photons. This may become
very important in the near future when gravitational waves from
distant sources will begin to be detected.

In summary, observable physics is invariant under disformal transformations,
 when applied to cosmology. We leave as future work the study of the 
appearance of false Ostrogradsky ghosts when the conformal or disformal 
factors include a kinetic term dependence, where hidden constraints 
should arise \citep{zumalacarregui2014transforming}.

\begin{acknowledgments}
G.D. and M.S. are grateful to N. Deruelle for fruitful discussions 
regarding mimetic gravity through disformal transformations. 
A.N. would like to thank the Yukawa Institute for Theoretical Physics
 at Kyoto University for warm hospitality where this work was advanced.
We would also like to thank Jinn-Ouk Gong and Masahide Yamaguchi
 for useful discussions especially during the workshop
 ``Miniworkshop on cosmology''.
This work was supported in part by the JSPS Research Fellowship
 for Young Scientists No.~263409 and in part by MEXT KAKENHI
No. 15H05888.
\end{acknowledgments}
\newpage
\appendix

\begin{flushleft}
\vspace{1cm}	
\bf APPENDICES
\end{flushleft}

\section{Inverse disformal transformation \label{sec:appendix}}
In the main text we considered a disformal transformation $g\to\bar{g}$
from the passive point of view. Namely we start from the barred frame
$\bar{g}=d\bar{s}^2$ and express the barred quantities in terms of 
the quantities in the unbarred frame $\bar{g}=\bar{g}(g)$.
Here we briefly discuss the inverse case where we start from a model 
in the unbarred frame with the metric $g$ and perform 
an inverse passive disformal transformation to work with $\bar{g}$. 
In other words, we begin with $g=ds^2$ and express the unbarred quantities
in terms of the barred one as $g=g(\bar{g})$. 

Let us fix that the barred metric is given by
\begin{align}
d\bar{s}^2=-dt^2+g_{ij}dx^idx^j,
\label{eq:metrica}
\end{align} 
namely we want to work with the proper cosmic time ($\bar{N}=1$) of 
$\bar{g}_{\mu\nu}$. By means of a disformal transformation,
\begin{align}
ds^2&=g_{\mu\nu} dx^\mu dx^\nu 
=(\bar{g}_{\mu\nu}-F(\phi,X)\phi_{,\mu}\phi_{,\nu}) dx^\mu dx^\nu\,,
\end{align}
 the unbarred line element reads
\begin{align}
ds^2
&=d\bar{s}^2+(1-\alpha^{-2}(t))dt^2=  -\alpha^{-2}(t) dt^2+g_{ij}dx^idx^j.
\label{eq:metricc}
\end{align}
This time, the suitable proper time redefinition in the unbarred frame is given by
\begin{align}
d\tilde{t}=\alpha^{-1}(t) dt\,
\label{eq:suit}
\end{align}
which leads us to
\begin{align}
ds^2=-d\tilde{t}^2+g_{ij}dx^idx^j.
\label{eq:metricd}
\end{align} 
In this form, the metric components are the same as those in the barred 
frame but with a rescaling of time $t\to\tilde{t}$. 
To summarize, starting from an action with the unbarred metric 
$g_{\mu\nu}$ and rewriting it in terms of the barred metric 
$\bar{g}_{\mu\nu}$ is equivalent to the rescaling of time from $t$ to 
$\tilde{t}$, i.e. $S[g(\bar{g});t]=S[\bar{g},\tilde{t}]$.

\section{Example of an observable: Redshift \label{sec:observables}}
Here we examine the frame independence of the measured redshift as a simple example of an observable quantity. The main point is that once we are given the action of a system, 
we can compute observable quantities in any frame and obtain the same result.

Let us assume that the background in the gravitational frame, where
the gravitational action is given by the Einstein-Hilbert one, is well 
described by a flat FLRW background,
\begin{align}
	ds^2=a^2(\eta)\left(-d\eta^2+d\mathbf{x}^2\right),
	\label{eq:graviframe}
\end{align}
where $a(\eta)$ is the scale factor as a function of the conformal time
defined by $d\eta=dt/a$. We assume the matter is coupled to
a disformally transformed metric, given by the transformation 
(\ref{eq:disdown}), which we call the matter frame metric.
It is expressed as 
\begin{align}
	d\bar{s}^2=-a^2(\eta)\alpha^2(\eta)d\eta^2+a^2(\eta)d\mathbf{x}^2
	=\bar{a}^2(\bar{\eta})\left(-d\bar{\eta}^2+d\mathbf{x}^2\right),
	\label{eq:matterframe}
\end{align}
where $\bar{a}(\bar{\eta})=a(\eta(\bar{\eta}))$ and in the last step we
used the time redefinition $d\bar{\eta}=\alpha d\eta$. 
Naturally, photons follow null geodesics of $\bar{g}_{\mu\nu}$, 
i.e. $d\bar{s}^2=0$. The energy of
a photon with four-momentum $\bar{k}^\mu=(\bar{k}^\eta,\bar{\bm{k}})$ 
measured by a comoving observer with four-momentum
$\bar{u}_\mu=(-\bar{a}(\bar{\eta}),0)$ is given by
\begin{align}
	\bar{\cal E}=-\bar{k}^\mu \bar{u}_\mu
	=\bar{a}(\bar{\eta})\bar{k}^{\bar{\eta}}.
\end{align}
One of the important points to be remembered when performing
a passive disformal transformation is that scalar quantities are 
invariant under such a transformation. This applies, in particular,
to the above energy measured by a comoving observer.
Alternatively, the same conclusion may be obtained by considering the 
transformation rules for a vector \eqref{eq:vectortransfup},
which implies $\bar{k}^\eta=\alpha^{-1}k^\eta$.
We can compute the energy in the unbarred frame as
\begin{align}
	{\cal E}=-{k}^\mu {u}_\mu=a(\eta) k^\eta.
\end{align}
Now since we have $\bar{k}^{\bar{\eta}}=\alpha \bar{k}^\eta=k^\eta$,
the measured energy is frame independent, $\bar{\cal E}={\cal E}$.
Consequently the redshift, which is defined as the ratio between the measured
photon energy at emission and observation,
\begin{align}
	1+z\equiv{\cal E}_{emit}/{\cal E}_{obs}=\frac{a(\eta_{obs})}{a(\eta_{emit})}
	=\frac{\bar{a}(\bar{\eta}_{obs})}{\bar{a}(\bar{\eta}_{emit})},
	\label{eq:redshift}
\end{align}
is frame independent as well. Once more the result is consistent 
with the fact that a cosmological pure disformal transformations is 
equivalent to a rescaling of the time. 

Finally, note that the physical speed of light is the same in both 
frames for a \textit{matter observer}, as long as matter and radiation 
are universally coupled to a single metric.
This can been easily seen from the condition $d\bar{s}^2=0$ 
which implies that the dispersion relation 
is $\bar{k}^{\bar{\eta}}=|\bar{\mathbf{k}}|=k^\eta$. 
We can further extend this results by noting that since the observed 
redshift is frame independent and the scale factor is not affected
by the disformal transformation, the luminosity distance is frame 
independent as well. This is in agreement with the work by 
Brax et al. \citep{brax2013cosmological} when matter
is universally coupled.  Other possible
 imprints in the Cosmological Microwave Background (CMB), such as a
 modification of the distribution function, are considered
 in \citep{van2013constraints}, where it is found that if matter and 
radiation are universally coupled to the disformal metric there is no 
observable difference in the CMB.

\bibliographystyle{jhep}
\bibliography{cosm_dis_inv_arxiv}

\providecommand{\href}[2]{#2}\begingroup\raggedright\begin{thebibliography}{10}

\bibitem{clifton2012modified}
T.~Clifton, P.~G. Ferreira, A.~Padilla, and C.~Skordis, {\it Modified gravity
  and cosmology},  {\em Physics Reports} {\bf 513} (2012), no.~1 1--189,
  [\href{http://arxiv.org/abs/1106.2476}{{\tt arXiv:1106.2476}}].

\bibitem{ade2015planck2}
P.~Ade, N.~Aghanim, M.~Arnaud, M.~Ashdown, J.~Aumont, C.~Baccigalupi,
  A.~Banday, R.~Barreiro, et~al., {\it Planck 2015 results. xiii. cosmological
  parameters},  \href{http://arxiv.org/abs/1502.01589}{{\tt arXiv:1502.01589}}.

\bibitem{ade2015planck3}
P.~Ade, N.~Aghanim, M.~Arnaud, M.~Ashdown, J.~Aumont, C.~Baccigalupi,
  A.~Banday, R.~Barreiro, N.~Bartolo, E.~Battaner, et~al., {\it Planck 2015
  results. xiv. dark energy and modified gravity},
  \href{http://arxiv.org/abs/1502.01590}{{\tt arXiv:1502.01590}}.

\bibitem{sahni2014model}
V.~Sahni, A.~Shafieloo, and A.~A. Starobinsky, {\it Model-independent evidence
  for dark energy evolution from baryon acoustic oscillations},  {\em The
  Astrophysical Journal Letters} {\bf 793} (2014), no.~2 L40,
  [\href{http://arxiv.org/abs/1406.2209}{{\tt arXiv:1406.2209}}].

\bibitem{ade2015planck}
P.~Ade, N.~Aghanim, M.~Arnaud, F.~Arroja, M.~Ashdown, J.~Aumont,
  C.~Baccigalupi, M.~Ballardini, A.~Banday, R.~Barreiro, et~al., {\it Planck
  2015. xx. constraints on inflation},
  \href{http://arxiv.org/abs/1502.02114}{{\tt arXiv:1502.02114}}.

\bibitem{brans1961mach}
C.~Brans and R.~H. Dicke, {\it Mach's principle and a relativistic theory of
  gravitation},  {\em Physical Review} {\bf 124} (1961), no.~3 925.

\bibitem{fujii2003scalar}
Y.~Fujii, K.-i. Maeda, et~al., {\em The scalar-tensor theory of gravitation}.
\newblock Cambridge University Press, 2003.

\bibitem{faraoni2004cosmology}
V.~Faraoni, {\em Cosmology in scalar-tensor gravity}, vol.~139.
\newblock Springer, 2004.

\bibitem{whitt1984fourth}
B.~Whitt, {\it Fourth-order gravity as general relativity plus matter},  {\em
  Physics Letters B} {\bf 145} (1984), no.~3 176--178.

\bibitem{jakubiec1988theories}
A.~Jakubiec and J.~Kijowski, {\it On theories of gravitation with nonlinear
  lagrangians},  {\em Physical Review D} {\bf 37} (1988), no.~6 1406.

\bibitem{kastrup2008advancements}
H.~A. Kastrup, {\it On the advancements of conformal transformations and their
  associated symmetries in geometry and theoretical physics},  {\em Annalen der
  Physik} {\bf 17} (2008), no.~9-10 631--690,
  [\href{http://arxiv.org/abs/0808.2730}{{\tt arXiv:0808.2730}}].

\bibitem{makino1991density}
N.~Makino and M.~Sasaki, {\it The density perturbation in the chaotic inflation
  with non-minimal coupling},  {\em Progress of Theoretical Physics} {\bf 86}
  (1991), no.~1 103--118.

\bibitem{faraoni2007pseudo}
V.~Faraoni and S.~Nadeau, {\it (pseudo) issue of the conformal frame
  revisited},  {\em Physical Review D} {\bf 75} (2007), no.~2 023501,
  [\href{http://arxiv.org/abs/gr-qc/0612075}{{\tt gr-qc/0612075}}].

\bibitem{deruelle2011conformal}
N.~Deruelle and M.~Sasaki, {\it Conformal equivalence in classical gravity: the
  example of "veiled" general relativity},  in {\em Cosmology, Quantum Vacuum
  and Zeta Functions}, pp.~247--260.
\newblock Springer, 2011.
\newblock \href{http://arxiv.org/abs/1007.3563}{{\tt arXiv:1007.3563}}.

\bibitem{gong2011conformal}
J.-O. Gong, J.-c. Hwang, W.~I. Park, M.~Sasaki, and Y.-S. Song, {\it Conformal
  invariance of curvature perturbation},  {\em Journal of Cosmology and
  Astroparticle Physics} {\bf 2011} (2011), no.~09 023,
  [\href{http://arxiv.org/abs/1107.1840}{{\tt arXiv:1107.1840}}].

\bibitem{white2012curvature}
J.~White, M.~Minamitsuji, and M.~Sasaki, {\it Curvature perturbation in
  multi-field inflation with non-minimal coupling},  {\em Journal of Cosmology
  and Astroparticle Physics} {\bf 2012} (2012), no.~07 039,
  [\href{http://arxiv.org/abs/1306.6186}{{\tt arXiv:1306.6186}}].

\bibitem{catena2007einstein}
R.~Catena, M.~Pietroni, and L.~Scarabello, {\it Einstein and jordan frames
  reconciled: A frame-invariant approach to scalar-tensor cosmology},  {\em
  Physical Review D} {\bf 76} (2007), no.~8 084039,
  [\href{http://arxiv.org/abs/astro-ph/0604492}{{\tt astro-ph/0604492}}].

\bibitem{chiba2013conformal}
T.~Chiba and M.~Yamaguchi, {\it Conformal-frame (in) dependence of cosmological
  observations in scalar-tensor theory},  {\em Journal of Cosmology and
  Astroparticle Physics} {\bf 2013} (2013), no.~10 040,
  [\href{http://arxiv.org/abs/1308.1142}{{\tt arXiv:1308.1142}}].

\bibitem{chiba2008extended}
T.~Chiba and M.~Yamaguchi, {\it Extended slow-roll conditions and rapid-roll
  conditions},  {\em Journal of Cosmology and Astroparticle Physics} {\bf 2008}
  (2008), no.~10 021, [\href{http://arxiv.org/abs/0807.4965}{{\tt
  arXiv:0807.4965}}].

\bibitem{jarv2014invariant}
L.~J\"arv, P.~Kuusk, M.~Saal, and O.~Vilson, {\it Invariant quantities in the
  scalar-tensor theories of gravitation},  {\em Phys. Rev. D} {\bf 91} (Jan,
  2015) 024041, [\href{http://arxiv.org/abs/1411.1947}{{\tt arXiv:1411.1947}}].

\bibitem{domenech2015conformal}
G.~Dom{\`e}nech and M.~Sasaki, {\it Conformal frame dependence of inflation},
  {\em Journal of Cosmology and Astroparticle Physics} {\bf 2015} (2015),
  no.~04 022, [\href{http://arxiv.org/abs/1501.07699}{{\tt arXiv:1501.07699}}].

\bibitem{capozziello2006cosmological}
S.~Capozziello, S.~Nojiri, S.~Odintsov, and A.~Troisi, {\it Cosmological
  viability of f (r)-gravity as an ideal fluid and its compatibility with a
  matter dominated phase},  {\em Physics Letters B} {\bf 639} (2006), no.~3
  135--143, [\href{http://arxiv.org/abs/astro-ph/0604431}{{\tt
  astro-ph/0604431}}].

\bibitem{nojiri2006modified}
S.~Nojiri and S.~D. Odintsov, {\it Modified f (r) gravity consistent with
  realistic cosmology: From a matter dominated epoch to a dark energy
  universe},  {\em Physical Review D} {\bf 74} (2006), no.~8 086005,
  [\href{http://arxiv.org/abs/hep-th/0608008}{{\tt hep-th/0608008}}].

\bibitem{briscese2007phantom}
F.~Briscese, E.~Elizalde, S.~Nojiri, and S.~Odintsov, {\it Phantom scalar dark
  energy as modified gravity: Understanding the origin of the big rip
  singularity},  {\em Physics Letters B} {\bf 646} (2007), no.~2 105--111,
  [\href{http://arxiv.org/abs/hep-th/0612220}{{\tt hep-th/0612220}}].

\bibitem{nojiri2011unified}
S.~Nojiri and S.~D. Odintsov, {\it Unified cosmic history in modified gravity:
  from f (r) theory to lorentz non-invariant models},  {\em Physics Reports}
  {\bf 505} (2011), no.~2 59--144, [\href{http://arxiv.org/abs/1011.0544}{{\tt
  arXiv:1011.0544}}].

\bibitem{bekenstein1993relation}
J.~D. Bekenstein, {\it Relation between physical and gravitational geometry},
  {\em Physical Review D} {\bf 48} (1993), no.~8 3641,
  [\href{http://arxiv.org/abs/gr-qc/9211017}{{\tt gr-qc/9211017}}].

\bibitem{brax2012shining}
P.~Brax, C.~Burrage, and A.-C. Davis, {\it Shining light on modifications of
  gravity},  {\em Journal of Cosmology and Astroparticle Physics} {\bf 2012}
  (2012), no.~10 016, [\href{http://arxiv.org/abs/1206.1809}{{\tt
  arXiv:1206.1809}}].

\bibitem{zumalacarregui2013dbi}
M.~Zumalac{\'a}rregui, T.~S. Koivisto, and D.~F. Mota, {\it Dbi galileons in
  the einstein frame: Local gravity and cosmology},  {\em Physical Review D}
  {\bf 87} (2013), no.~8 083010, [\href{http://arxiv.org/abs/1210.8016}{{\tt
  arXiv:1210.8016}}].

\bibitem{kaloper2004disformal}
N.~Kaloper, {\it Disformal inflation},  {\em Physics Letters B} {\bf 583}
  (2004), no.~1 1--13, [\href{http://arxiv.org/abs/hep-ph/0312002}{{\tt
  hep-ph/0312002}}].

\bibitem{koivisto2008disformal}
T.~S. Koivisto, {\it Disformal quintessence},
  \href{http://arxiv.org/abs/0811.1957}{{\tt arXiv:0811.1957}}.

\bibitem{zumalacarregui2010disformal}
M.~Zumalacarregui, T.~Koivisto, D.~Mota, and P.~Ruiz-Lapuente, {\it Disformal
  scalar fields and the dark sector of the universe},  {\em Journal of
  Cosmology and Astroparticle Physics} {\bf 2010} (2010), no.~05 038,
  [\href{http://arxiv.org/abs/1004.2684}{{\tt arXiv:1004.2684}}].

\bibitem{van2015disformal}
C.~van~de Bruck and J.~Morrice, {\it {Disformal couplings and the dark sector
  of the universe}},  {\em Journal of Cosmology and Astroparticle Physics} {\bf
  1504} (2015), no.~04 036, [\href{http://arxiv.org/abs/1501.03073}{{\tt
  arXiv:1501.03073}}].

\bibitem{clayton2001scalar}
M.~Clayton and J.~Moffat, {\it A scalar-tensor cosmological model with
  dynamical light velocity},  {\em Physics Letters B} {\bf 506} (2001), no.~1
  177--186, [\href{http://arxiv.org/abs/gr-qc/0101126}{{\tt gr-qc/0101126}}].

\bibitem{moffat2003bimetric}
J.~Moffat, {\it Bimetric gravity theory, varying speed of light and the dimming
  of supernovae},  {\em International Journal of Modern Physics D} {\bf 12}
  (2003), no.~02 281--298, [\href{http://arxiv.org/abs/gr-qc/0202012}{{\tt
  gr-qc/0202012}}].

\bibitem{magueijo2003new}
J.~Magueijo, {\it New varying speed of light theories},  {\em Reports on
  Progress in Physics} {\bf 66} (2003), no.~11 2025,
  [\href{http://arxiv.org/abs/astro-ph/0305457}{{\tt astro-ph/0305457}}].

\bibitem{magueijo2009bimetric}
J.~Magueijo, {\it Bimetric varying speed of light theories and primordial
  fluctuations},  {\em Physical Review D} {\bf 79} (2009), no.~4 043525,
  [\href{http://arxiv.org/abs/0807.1689}{{\tt arXiv:0807.1689}}].

\bibitem{brax2014explaining}
P.~Brax and C.~Burrage, {\it {Explaining the Proton Radius Puzzle with
  Disformal Scalars}},  {\em Phys.Rev.} {\bf D91} (2015), no.~4 043515,
  [\href{http://arxiv.org/abs/1407.2376}{{\tt arXiv:1407.2376}}].

\bibitem{deruelle2014disformal}
N.~Deruelle and J.~Rua, {\it Disformal transformations, veiled general
  relativity and mimetic gravity},  {\em Journal of Cosmology and Astroparticle
  Physics} {\bf 2014} (2014), no.~09 002,
  [\href{http://arxiv.org/abs/1407.0825}{{\tt arXiv:1407.0825}}].

\bibitem{sakstein2014disformal}
J.~Sakstein, {\it Disformal theories of gravity: from the solar system to
  cosmology},  {\em Journal of Cosmology and Astroparticle Physics} {\bf 2014}
  (2014), no.~12 012, [\href{http://arxiv.org/abs/1409.1734}{{\tt
  arXiv:1409.1734}}].

\bibitem{sakstein2015towards}
J.~Sakstein, {\it Towards viable cosmological models of disformal theories of
  gravity},  {\em Physical Review D} {\bf 91} (2015), no.~2 024036,
  [\href{http://arxiv.org/abs/1409.7296}{{\tt arXiv:1409.7296}}].

\bibitem{koivisto2012screening}
T.~S. Koivisto, D.~F. Mota, and M.~Zumalac{\'a}rregui, {\it Screening
  modifications of gravity through disformally coupled fields},  {\em Physical
  review letters} {\bf 109} (2012), no.~24 241102,
  [\href{http://arxiv.org/abs/1205.3167}{{\tt arXiv:1205.3167}}].

\bibitem{van2013constraints}
C.~van~de Bruck, J.~Morrice, and S.~Vu, {\it Constraints on nonconformal
  couplings from the properties of the cosmic microwave background radiation},
  {\em Physical review letters} {\bf 111} (2013), no.~16 161302,
  [\href{http://arxiv.org/abs/1303.1773}{{\tt arXiv:1303.1773}}].

\bibitem{brax2013cosmological}
P.~Brax, C.~Burrage, A.-C. Davis, and G.~Gubitosi, {\it Cosmological tests of
  the disformal coupling to radiation},  {\em Journal of Cosmology and
  Astroparticle Physics} {\bf 2013} (2013), no.~11 001,
  [\href{http://arxiv.org/abs/1306.4168}{{\tt arXiv:1306.4168}}].

\bibitem{brax2014constraining}
P.~Brax and C.~Burrage, {\it Constraining disformally coupled scalar fields},
  {\em Physical Review D} {\bf 90} (2014), no.~10 104009,
  [\href{http://arxiv.org/abs/1407.1861}{{\tt arXiv:1407.1861}}].

\bibitem{horndeski1974second}
G.~W. Horndeski, {\it Second-order scalar-tensor field equations in a
  four-dimensional space},  {\em International Journal of Theoretical Physics}
  {\bf 10} (1974), no.~6 363--384.

\bibitem{Deffayet:2011gz}
C.~Deffayet, X.~Gao, D.~Steer, and G.~Zahariade, {\it {From k-essence to
  generalised Galileons}},  {\em Phys.Rev.} {\bf D84} (2011) 064039,
  [\href{http://arxiv.org/abs/1103.3260}{{\tt arXiv:1103.3260}}].

\bibitem{kobayashi2011generalized}
T.~Kobayashi, M.~Yamaguchi, and J.~Yokoyama, {\it Generalized g-inflation with
  the most general second-order field equations},  {\em Progress of Theoretical
  Physics} {\bf 126} (2011), no.~3 511--529,
  [\href{http://arxiv.org/abs/1105.5723}{{\tt arXiv:1105.5723}}].

\bibitem{gleyzes2013essential}
J.~Gleyzes, D.~Langlois, F.~Piazza, and F.~Vernizzi, {\it Essential building
  blocks of dark energy},  {\em Journal of Cosmology and Astroparticle Physics}
  {\bf 2013} (2013), no.~08 025, [\href{http://arxiv.org/abs/1304.4840}{{\tt
  arXiv:1304.4840}}].

\bibitem{gleyzes2014healthy}
J.~Gleyzes, D.~Langlois, F.~Piazza, and F.~Vernizzi, {\it {Healthy theories
  beyond Horndeski}},  {\em Phys. Rev. Lett.} {\bf 114} (2015), no.~21 211101,
  [\href{http://arxiv.org/abs/1404.6495}{{\tt arXiv:1404.6495}}].

\bibitem{gleyzes2014exploring}
J.~Gleyzes, D.~Langlois, F.~Piazza, , and F.~Vernizzi, {\it Exploring
  gravitational theories beyond horndeski},  {\em Journal of Cosmology and
  Astroparticle Physics} {\bf 2015} (2015), no.~02 018,
  [\href{http://arxiv.org/abs/1408.1952}{{\tt arXiv:1408.1952}}].

\bibitem{gleyzes2014unifying}
J.~Gleyzes, D.~Langlois, and F.~Vernizzi, {\it A unifying description of dark
  energy},  {\em International Journal of Modern Physics D} {\bf 23} (2014),
  no.~13 1443010, [\href{http://arxiv.org/abs/1411.3712}{{\tt
  arXiv:1411.3712}}].

\bibitem{Gao:2014soa}
X.~Gao, {\it {Unifying framework for scalar-tensor theories of gravity}},  {\em
  Phys.Rev.} {\bf D90} (2014), no.~8 081501,
  [\href{http://arxiv.org/abs/1406.0822}{{\tt arXiv:1406.0822}}].

\bibitem{zumalacarregui2014transforming}
M.~Zumalac{\'a}rregui and J.~Garc{\'\i}a-Bellido, {\it Transforming gravity:
  from derivative couplings to matter to second-order scalar-tensor theories
  beyond the horndeski lagrangian},  {\em Physical Review D} {\bf 89} (2014),
  no.~6 064046, [\href{http://arxiv.org/abs/1308.4685}{{\tt arXiv:1308.4685}}].

\bibitem{bettoni2013disformal}
D.~Bettoni and S.~Liberati, {\it Disformal invariance of second order
  scalar-tensor theories: Framing the horndeski action},  {\em Physical Review
  D} {\bf 88} (2013), no.~8 084020, [\href{http://arxiv.org/abs/1306.6724}{{\tt
  arXiv:1306.6724}}].

\bibitem{bettoni2015shaken}
D.~Bettoni and M.~Zumalacárregui, {\it {Kinetic mixing in scalar-tensor
  theories of gravity}},  {\em Phys. Rev.} {\bf D91} (2015) 104009,
  [\href{http://arxiv.org/abs/1502.02666}{{\tt arXiv:1502.02666}}].

\bibitem{creminelli2014resilience}
P.~Creminelli, J.~Gleyzes, J.~Nore{\~n}a, and F.~Vernizzi, {\it Resilience of
  the standard predictions for primordial tensor modes},  {\em Physical review
  letters} {\bf 113} (2014), no.~23 231301,
  [\href{http://arxiv.org/abs/1407.8439}{{\tt arXiv:1407.8439}}].

\bibitem{minamitsuji2014disformal}
M.~Minamitsuji, {\it Disformal transformation of cosmological perturbations},
  {\em Physics letters B} {\bf 737} (2014) 139--150,
  [\href{http://arxiv.org/abs/1409.1566}{{\tt arXiv:1409.1566}}].

\bibitem{tsujikawa2014disformal}
S.~Tsujikawa, {\it {Disformal invariance of cosmological perturbations in a
  generalized class of Horndeski theories}},  {\em Journal of Cosmology and
  Astroparticle Physics} {\bf 1504} (2015) 043,
  [\href{http://arxiv.org/abs/1412.6210}{{\tt arXiv:1412.6210}}].

\bibitem{watanabe2015multi}
Y.~Watanabe, A.~Naruko, and M.~Sasaki, {\it {Multi-disformal invariance of
  non-linear primordial perturbations}},  {\em Europhys. Lett.} {\bf 111}
  (2015) 39002, [\href{http://arxiv.org/abs/1504.00672}{{\tt
  arXiv:1504.00672}}].

\bibitem{motohashi2015disformal}
H.~Motohashi and J.~White, {\it Disformal invariance of curvature
  perturbation},  \href{http://arxiv.org/abs/1504.00846}{{\tt
  arXiv:1504.00846}}.

\bibitem{gleyzes2015effective}
J.~Gleyzes, D.~Langlois, M.~Mancarella, and F.~Vernizzi, {\it {Effective Theory
  of Interacting Dark Energy}},  {\em Journal of Cosmology and Astroparticle
  Physics} {\bf 1508} (2015), no.~08 054,
  [\href{http://arxiv.org/abs/1504.05481}{{\tt arXiv:1504.05481}}].

\bibitem{chamseddine2014cosmology}
A.~H. Chamseddine, V.~Mukhanov, and A.~Vikman, {\it Cosmology with mimetic
  matter},  {\em Journal of Cosmology and Astroparticle Physics} {\bf 2014}
  (2014), no.~06 017, [\href{http://arxiv.org/abs/1403.3961}{{\tt
  arXiv:1403.3961}}].

\bibitem{chamseddine2013mimetic}
A.~H. Chamseddine and V.~Mukhanov, {\it Mimetic dark matter},  {\em Journal of
  High Energy Physics} {\bf 2013} (2013), no.~11 1--5,
  [\href{http://arxiv.org/abs/1308.5410}{{\tt arXiv:1308.5410}}].

\bibitem{ellis2005c}
G.~F. Ellis and J.-P. Uzan, {\it c is the speed of light, isn't it?},  {\em
  American journal of physics} {\bf 73} (2005), no.~3 240--247,
  [\href{http://arxiv.org/abs/gr-qc/0305099}{{\tt gr-qc/0305099}}].

\bibitem{ellis2007note}
G.~F. Ellis, {\it Note on varying speed of light cosmologies},  {\em General
  Relativity and Gravitation} {\bf 39} (2007), no.~4 511--520,
  [\href{http://arxiv.org/abs/astro-ph/0703751}{{\tt astro-ph/0703751}}].

\bibitem{maldacena2003non}
J.~Maldacena, {\it Non-gaussian features of primordial fluctuations in single
  field inflationary models},  {\em Journal of High Energy Physics} {\bf 2003}
  (2003), no.~05 013, [\href{http://arxiv.org/abs/astro-ph/0210603}{{\tt
  astro-ph/0210603}}].

\bibitem{wald2010general}
R.~M. Wald, {\em General relativity}.
\newblock University of Chicago press, 2010.

\bibitem{birrell1984quantum}
N.~D. Birrell and P.~C.~W. Davies, {\em Quantum fields in curved space}.
\newblock Cambridge university press, 1984.

\end{thebibliography}\endgroup

\end{document}